\documentclass[superscriptaddress,altaffilletter,amsmath,amssymb,aps,twocolumn,prl,nobibnotes,nofootinbib]{revtex4-2}

\usepackage[english]{babel}
\usepackage[colorinlistoftodos]{todonotes}
\usepackage{multirow}
\usepackage{comment}
\usepackage{relsize}
\usepackage{float}
\usepackage{textcomp}
\usepackage[colorlinks=true,allcolors=.,urlcolor=blue]{hyperref}
\usepackage{lineno}
\usepackage[justification=justified]{subcaption}
\usepackage[justification=justified, figurename=FIG.]{caption}

\DeclareCaptionLabelFormat{UC}{\MakeUppercase{#1} #2}
\usepackage{xcolor}
\usepackage{graphicx}
\usepackage{subcaption}
\usepackage{epstopdf}
\usepackage{color}
\usepackage{dcolumn}
\usepackage{bm}
\usepackage{hyperref}
\usepackage{booktabs} 
\usepackage{siunitx}
\usepackage{placeins}
\usepackage{xspace}
\usepackage{threeparttable}

\usepackage{comment}

\newcommand\Tstrut{\rule{0pt}{2.6ex}}         

\begin{document}

\title{Observation of low-lying isomeric states in $^{136}$Cs: a new avenue for dark matter and solar neutrino detection in xenon detectors}

\author{S.J.~Haselschwardt} \email[]{scotthaselschwardt@lbl.gov}
\affiliation{Lawrence Berkeley National Laboratory, 1 Cyclotron Road, Berkeley, CA 94720, USA}

\author{B.G.~Lenardo} \email[Corresponding author: ]{blenardo@slac.stanford.edu}
\affiliation{SLAC National Accelerator Laboratory, 2575 Sand Hill Rd, Menlo Park, CA 94025, USA}

\author{T.~Daniels}
\affiliation{Department of Physics and Physical Oceanography, University of North Carolina at Wilmington, Wilmington, NC 28403, USA}

\author{S.W.~Finch}
\affiliation{Department of Physics, Duke University, and Triangle Universities Nuclear Laboratory (TUNL), Durham, NC 27708, USA}

\author{F.Q.L.~Friesen}
\affiliation{Department of Physics, Duke University, and Triangle Universities Nuclear Laboratory (TUNL), Durham, NC 27708, USA}

\author{C.R.~Howell}
\affiliation{Department of Physics, Duke University, and Triangle Universities Nuclear Laboratory (TUNL), Durham, NC 27708, USA}

\author{C.R.~Malone} \thanks{Present address: Savannah River National Laboratory,  Aiken, SC 29802, USA}
\affiliation{Department of Physics, Duke University, and Triangle Universities Nuclear Laboratory (TUNL), Durham, NC 27708, USA}

\author{E.~Mancil}
\affiliation{Department of Physics, Duke University, and Triangle Universities Nuclear Laboratory (TUNL), Durham, NC 27708, USA}

\author{W.~Tornow}
\affiliation{Department of Physics, Duke University, and Triangle Universities Nuclear Laboratory (TUNL), Durham, NC 27708, USA}

\date{\today}
\begin{abstract}
\noindent

We report on new measurements establishing the existence of low-lying isomeric states in $^{136}$Cs using $\gamma$ rays produced in $^{136}$Xe(p,n)$^{136}$Cs reactions. Two states with $\mathcal{O}(100)$~ns lifetimes are placed in the decay sequence of the $^{136}$Cs levels that are populated in charged-current interactions of solar neutrinos and fermionic dark matter with $^{136}$Xe. Xenon-based experiments can therefore exploit a delayed-coincidence tag of these interactions, greatly suppressing backgrounds to enable spectroscopic studies of solar neutrinos and dark matter.
\end{abstract}

\maketitle


\emph{Introduction --} Future xenon-based experiments searching for dark matter and neutrinoless double-beta decay ($0\nu\beta\beta$) will deploy roughly 5--100~t of target mass. 
These detectors will provide powerful searches for rare interactions owing to their large targets, extremely low intrinsic backgrounds, and event reconstruction capabilities. 
In this work we present measurements which make possible the identification of a new class of events in these detectors -- charged-current (CC) interactions on $^{136}$Xe nuclei -- which can enable novel studies of low-energy solar neutrinos and provide unprecedented sensitivity to certain models of dark matter.

The CC ``neutrino capture" process, famously exploited by Ray Davis in the first detection of solar neutrinos~\cite{Davis:1968cp}, will be observable in Xe detectors as $\nu_e + \,^{A}_{54}\textrm{Xe} \to \,^{A}_{55}\textrm{Cs}^{(*)} + e^{-}$, where a Xe nucleus is converted into a (possibly excited) Cs nucleus via a Gamow-Teller transition ($\Delta J = 1$). The signal generated in the detector is the combination of the outgoing electron and any $\gamma$ rays/conversion electrons emitted as the Cs nucleus relaxes to its ground state. 
Measuring the energies of the final state particles provides complete reconstruction of the neutrino's kinetic energy, in contrast to the elastic scattering of neutrinos on electrons or nuclei~\cite{Dutta:2019oaj,Newstead:2018muu,Baudis:2013bba}.  
The total energy deposited in the detector
equals the neutrino energy minus the reaction threshold $Q$ (the mass difference between the Xe and Cs isobars). If $Q$ is low enough, this reaction can be used to measure neutrinos from the solar carbon-nitrogen-oxygen (CNO) cycle, an elusive signal which has only been observed in one experiment to date~\cite{BOREXINO:2020aww,BOREXINO:2022abl} and which plays a crucial role in determining the solar metallicity. This reaction can also provide a unique measurement of $^{7}$Be neutrinos, which may enable novel measurements of the solar core temperature~\cite{Bahcall:1994cf}. 
A variety of targets and techniques have been proposed to detect the lower-energy solar neutrino components in this way~\cite{Raghavan:1976yc,Raghavan:1997ad,Ejiri:1999rk,Zuber:2002wi,Wang:2020tdm, Chavarria:2021rbw}; however, none have been realized at scale.

The same final state can be used to search for models of fermionic dark matter in which the dark matter particle $\chi$ carries lepton number and interacts with Standard Model particles via a right-handed gauge boson $W'$~\cite{Dror:2019onn,Dror:2019dib}. CC interactions of $\chi$ with Xe nuclei, $\chi + \,^{A}_{54}\textrm{Xe} \to \,^{A}_{55}\textrm{Cs}^{(*)} + e^{-}$, convert the mass energy of $\chi$ into the mass and kinetic energy of the outgoing electron and excitation of the Cs nucleus. A recent search for these signals in a 200~kg liquid xenon (LXe) time projection chamber (TPC) produced constraints that are competitive with state-of-the-art collider searches~\cite{EXO-200:2022adi}.

Xenon-based detectors may be uniquely positioned to perform world-class measurements of CC interactions~\cite{Haselschwardt:2020ffr}.
The $\beta\beta$ isotope $^{136}$Xe is particularly well-suited as a CC reaction target~\cite{Raghavan:1997ad}: it features a low threshold of $Q = 90.3$~keV and relatively large cross section due to the sizable Gamow-Teller transition strengths~\cite{Puppe:2011zz,Frekers:2013} connecting the $0^+$ $^{136}$Xe ground state and the lowest-lying $1^+$ excited states of $^{136}$Cs near 590~keV and 850~keV. 
Using shell-model (SM) calculations, Ref.~\cite{Haselschwardt:2020ffr} predicts that the $^{136}$Cs $1^+$ states relax through an isomeric state, setting up a delayed-coincidence signature that would allow unambiguous identification of CC interactions and enable background-free measurements of these signals in current- and next-generation detectors. This opportunity depends critically on the level structure of $^{136}$Cs below $\sim$590~keV and particularly on the unmeasured $\gamma$-ray emission properties of the lowest-lying states.

Until recently the low-lying level structure of $^{136}$Cs has been relatively unexplored~\cite{McCutchan:2018}. Current data stems mainly from two experimental campaigns focused on providing inputs for the calculation of nuclear matrix elements for $\beta\beta$ decay of $^{136}$Xe. 
First, the high-resolution $^{136}\textrm{Xe}(^{3}\textrm{He},t)$ measurements reported in Ref.~\cite{Puppe:2011zz} established the spectrum of $1^+$ states and the Gamow-Teller transition strengths to these states from the $0^+$ $^{136}$Xe ground state.
Second, a campaign using the $^{138}$Ba(d,$\alpha$) reaction reported several new low-lying states in $^{136}$Cs~\cite{Rebeiro:2016pvi,Rebeiro:2018svw,RebeiroThesis,capetown_placeholder}. 
There is only one study of excitations near the ground state in $^{136}$Cs using $\gamma$ rays, which established a single $4^+$ level at 104.8(3)~keV relevant for CC events~\cite{Wimmer:2011}.

In this paper we report on an experiment to characterize the energies and lifetimes of the low-lying states in $^{136}$Cs using $\gamma$ rays produced in (p,n) reactions on $^{136}$Xe. We identify many new nuclear transitions, several of which have $\mathcal{O}(100)$~ns lifetimes. 
Using $\gamma$-$\gamma$ coincidences we reconstruct a level scheme which describes the decay of the $1^{+}$ states of interest and identify isomeric states which will produce the delayed-coincidence signature required for low-background study of CC neutrino and dark matter interactions in xenon-based experiments.

\emph{Data collection and analysis --} Measurements were performed at the Triangle Universities Nuclear Laboratory. A pulsed beam of 7~MeV protons with period of 1.6~$\mu$s and pulse width of 2~ns was directed through a target cell containing Xe gas enriched to 94\% in $^{136}$Xe. The beam current was kept between 5--15~nA throughout the run, giving a Xe(p,n)Cs reaction rate of $\sim$1/pulse. The target was a 1.3~cm tall, 1.6~cm diameter cylinder with 25~$\mu$m thick polyethylene naphthalate (PEN) films serving as beam entrance and exit windows. The cell was housed in a cylindrical aluminum vacuum chamber with inner diameter 30.5~cm and a 6.4~mm wall thickness. The inner wall of the chamber was lined with 1~mm of lead, with gaps through which $\gamma$-ray detectors viewed the target. 

Gamma rays were measured by four high-purity germanium (HPGe) detectors placed outside the vacuum chamber: two 60\% relative efficiency coaxial detectors, which provided high detection efficiency for $\gamma$ rays between 50--3000~keV, and two planar low-energy photon spectrometers (LEPS), which provided $\mathcal{O}(10)$~ns time resolution and sensitivity between 20--600~keV. To ensure high-quality measurements of $\gamma$~rays below 100~keV, one LEPS detector viewed the target through a 127~$\mu$m Kapton window and was positioned 51.3~cm from the target center. All other detectors viewed the target through the wall of the target chamber at a distance of $25\pm2$~cm.
Detectors were shielded from the beam dump and beamline components upstream of the target using tungsten and lead. A Mesytec MDPP-16 digitizer was used to record the energy and time for each detected $\gamma$ ray. The time reference was provided by a capacitive sensor through which the proton beam passed before entering the target chamber. 

The energy scale and efficiency of each detector were calibrated using a combination of standard $^{133}$Ba, $^{137}$Cs, $^{60}$Co $\gamma$-ray sources, a mixed source containing $^{241}$Am, $^{57}$Co, $^{54}$Mn, $^{65}$Zn, and known transitions from the $\beta^-$ decay of $^{136}$Cs in the target.\footnote{During this procedure, it was discovered that the 66.881 keV $\gamma$ from the ground state $\beta^-$ decay of $^{136}$Cs has an incorrectly assigned intensity in the present evaluation~\cite{McCutchan:2018}. This evaluation provides $I_{\gamma} = 4.79(20)$, whereas our data prefer $I_{\gamma}=12.5(1)$ as reported in Ref.~\cite{Griffioen1975}.} During the course of data taking the energy scale varied by at most 0.8\%, and run-by-run corrections are applied in the analysis.
The LEPS detectors' timing capabilities were calibrated using the 308~keV isomeric transition in $^{48}$V  ($\tau = 10.26\pm0.06$~ns~\cite{Chen:2022}) in a dedicated run of $^{48}$Ti(p,n) reactions on a $^{\rm nat}$Ti foil target. We measure a lifetime of $10.6\pm0.4$~ns, in good agreement with the accepted value.

\begin{figure}[t]
\center
\includegraphics[width=\columnwidth]{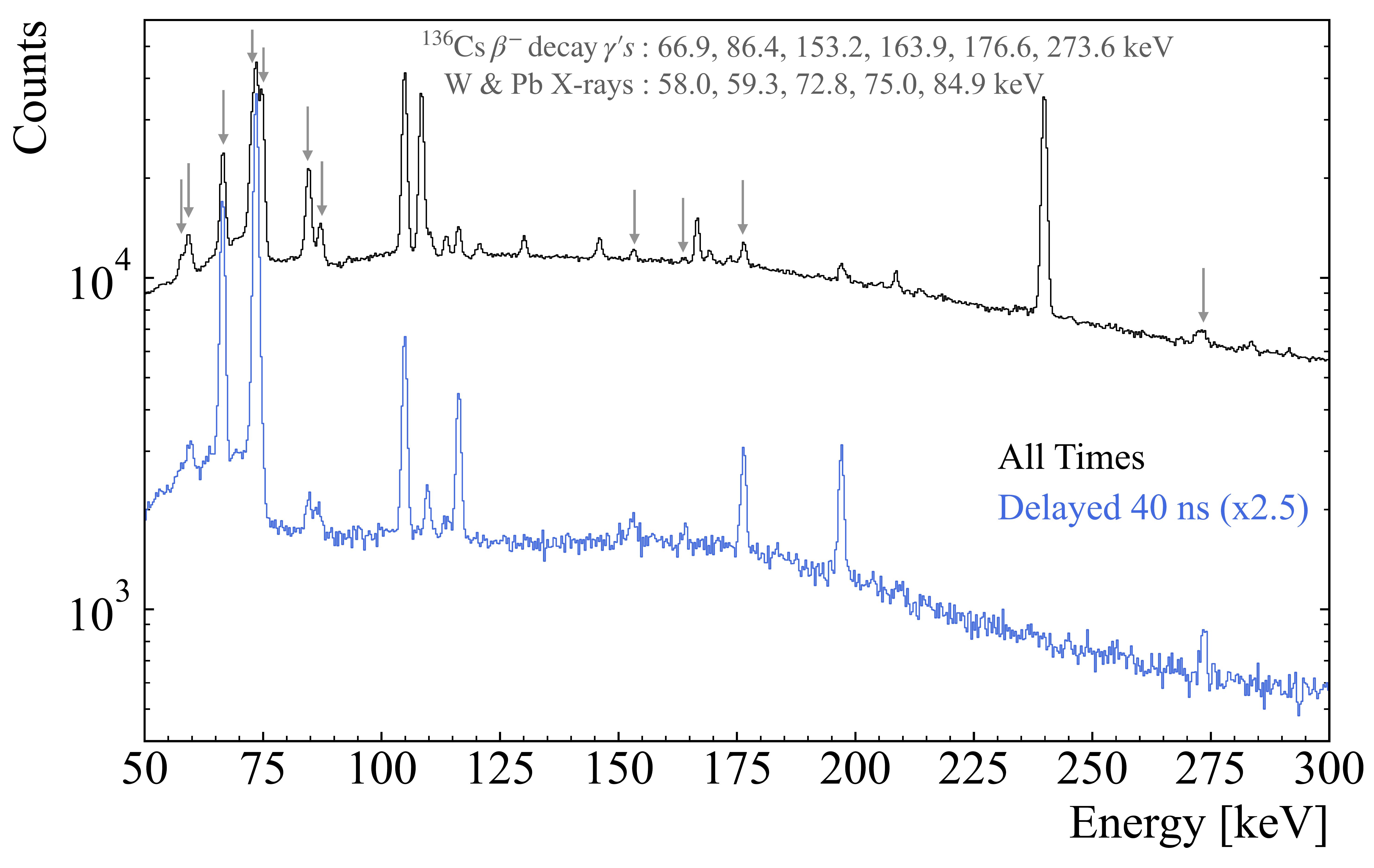}
\caption{Energy spectrum of events in the near LEPS detector. The black curve shows the spectrum with no selection on arrival time. The blue curve displays the delayed spectrum, defined as events arriving $>40$~ns after the beam pulse. Grey arrows indicate background lines from $^{136}$Cs $\beta^{-}$ decay and X-rays from tungsten and lead shielding.}
\label{fig:singles}
\end{figure}

\emph{Results --} A typical energy spectrum of single-hit events in a LEPS detector is shown in Fig.~\ref{fig:singles}. We set an analysis threshold at 50~keV, below which the spectrum is dominated by background from proton-induced X-ray emission in the Xe target. 
We identify 65 $\gamma$-ray lines between 50--1100~keV from previously unobserved transitions in $^{136}$Cs and measure their energies with an uncertainty of 0.1~keV. A complete list is available in the Supplementary Material. These signals can be classified into ``prompt'' -- defined here to be within 40~ns of the beam pulse -- or ``delayed''. Eight of the new transitions are delayed, featuring time distributions which decay over periods of $\mathcal{O}(100)$~ns. These, along with steady-state backgrounds from target activation, appear in the delayed spectrum in Fig.~\ref{fig:singles}. In addition to the transitions reported here for the first time, we observe the known lines at 104.8(3) and 517.9(1)~keV associated with direct transitions to the ground state~\cite{Wimmer:2011}, with the latter appearing constant in time due to the long lifetime of the parent $8^-$ level ($\tau=25.2(3)$~s). 

\begin{figure}[tb]
\center
\includegraphics[width=\columnwidth]{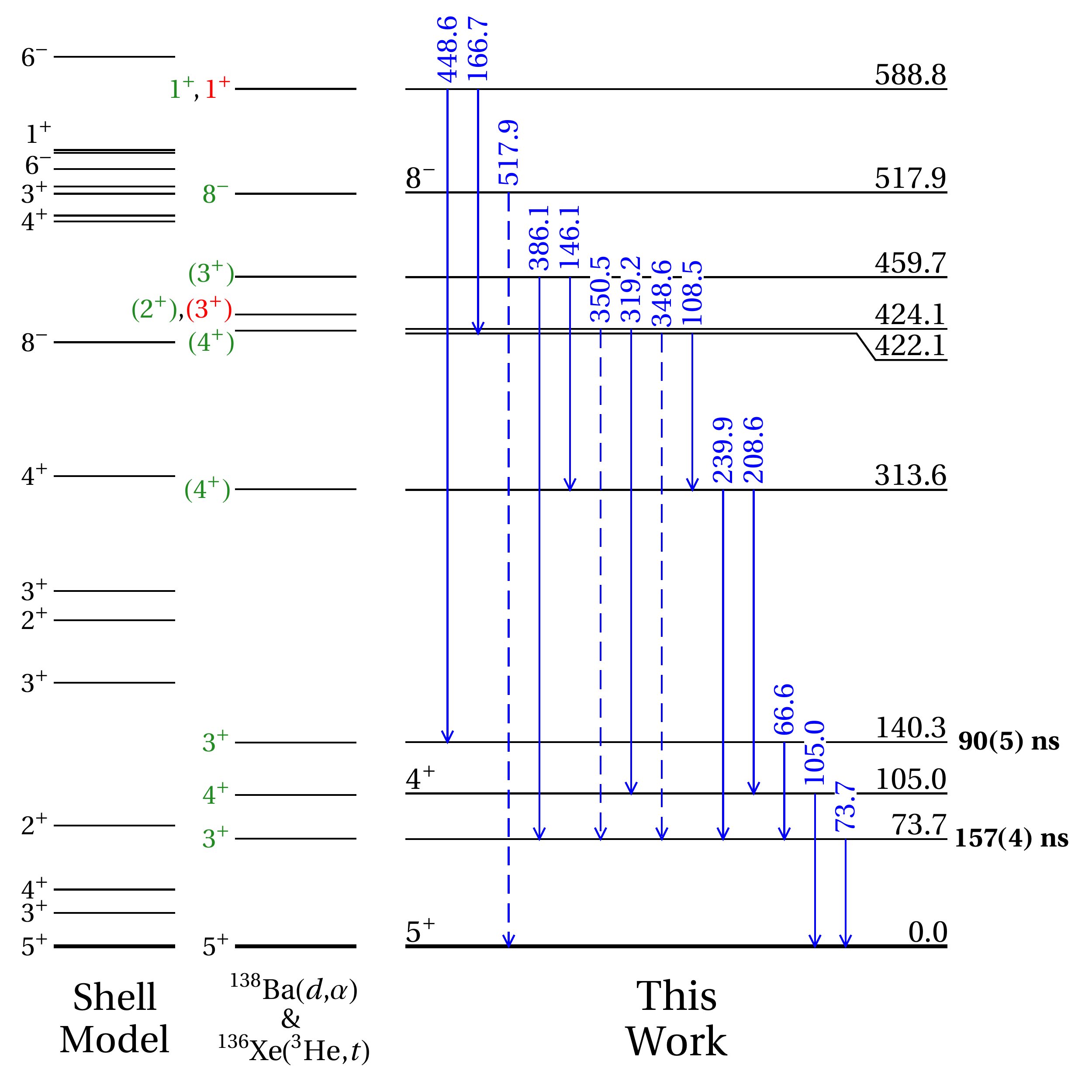}
\caption{
The rightmost column shows our proposed level scheme for $^{136}$Cs up to 590~keV, with energies given in keV and measured state lifetimes. A scheme using all observed coincidences is given in the Supplemental Material. Blue lines show measured $\gamma$-ray transitions, where those drawn as dashed lines have weak coincidence intensity but are observed in the singles spectrum. The left column shows the spectrum of states predicted by the shell model of Ref.~\cite{Haselschwardt:2020ffr}. The central column shows levels measured using $^{138}$Ba(d,$\alpha$)~\cite{capetown_placeholder} and $^{136}$Xe($^3$He,t)~\cite{Puppe:2011zz,Frekers:2013} reactions with their assigned $J^{\pi}$ values shown in green and red, respectively.
}
\label{fig:scheme}
\end{figure}

The low-lying level scheme of $^{136}$Cs is constructed through analysis of $\gamma$-$\gamma$ coincidences. A 1~$\mu$s window is used to construct two-dimensional coincidence matrices of the energy in each detector. Accidental coincidences measured \emph{in situ} are subtracted off. The resulting matrix and associated coincidence gates are analyzed using the RadWare program~\cite{Radford:1994vy}. 

The proposed level scheme is shown in Fig.~\ref{fig:scheme}. For comparison, we show the SM predictions from Ref.~\cite{Haselschwardt:2020ffr} as well as the spectrum measured using $^{138}$Ba(d,$\alpha$)~\cite{capetown_placeholder} and $^{136}$Xe($^3$He,t) reactions~\cite{Puppe:2011zz,Frekers:2013}. Nearly all of our reconstructed levels can be mapped to levels observed in previous experiments. However, the high-resolution $\gamma$-ray measurements in this work enable us to reduce the uncertainties in their energy by an order of magnitude. In addition, we reconstruct a doublet in the vicinity of 420~keV which is unresolved in charged-particle-based experiments and reported here for the first time. One of these two states plays a crucial role in CC interactions, as discussed below.

\begin{figure}[t]
\center
\includegraphics[width=\columnwidth]{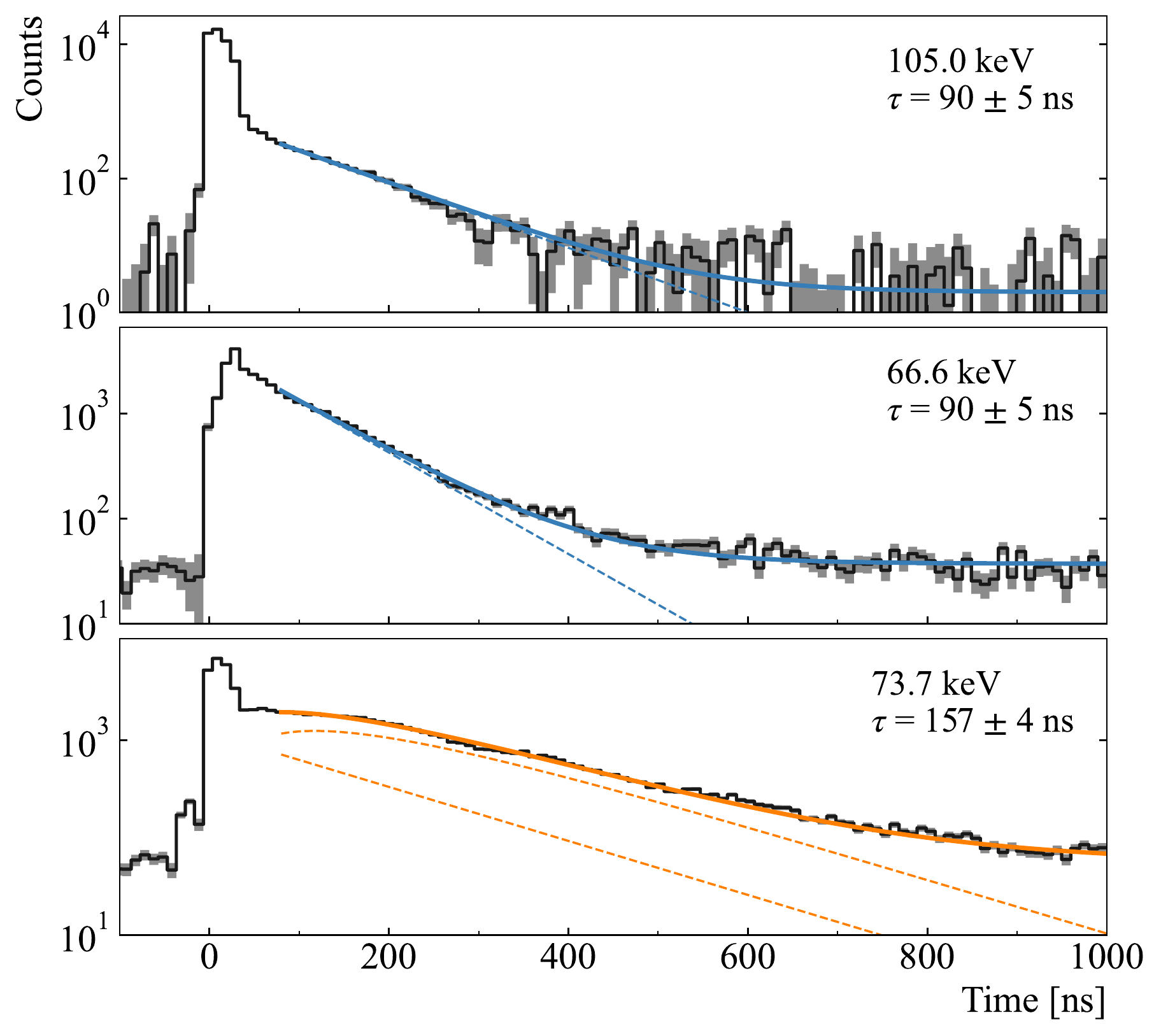}
\caption{Background-subtracted $\gamma$-ray arrival times of the three delayed transitions in our level scheme. The 105~keV and 67~keV $\gamma$'s are fit with exponential and constant background terms only, while the fit to the 74~keV transition contains a term accounting for the additional delay caused by preceding 67~keV transitions, in accordance with the proposed level scheme. X-rays from lead produce the prompt peak observed in the 74~keV distribution.
}
\label{fig:timing}
\end{figure}

Of the eight delayed transitions, those at 66.6, 73.7, and 105.0~keV are measured with sufficient strength in $\gamma$-$\gamma$ coincidences to be placed in our level scheme. Their background-subtracted time distributions are shown in Fig.~\ref{fig:timing}. The delayed components of the 66.6~keV and 105.0~keV transitions are well described by a single exponential distribution with the same lifetime. The 105.0~keV component also contains a significant prompt component. 
This indicates that the $105.0 \rightarrow \mathrm{G.S.}$ decay itself is prompt, and that the observed exponential distribution comes from the isomeric 140.3~keV state feeding the 105.0~keV level through a 35.3~keV transition which falls below our analysis threshold. 
Selecting only 105.0~keV $\gamma$'s that occur in coincidence with the 319~keV transition (thereby skipping the 140.3~keV level) produces a time distribution with no delayed component, supporting this conclusion. The 73.7~keV transition is placed between the lowest-lying excited state and the ground state. As such, it can either be fed by the long-lived 140.3~keV state or by other (prompt) transitions. Components to model each of these scenarios are included in the fit to its time distribution. The fit lifetime for the 73.7~keV state is $\tau = 157\pm4$~ns and is the longest observed in our experiment.

From the measured relative $\gamma$-ray intensities we conclude that approximately 99\% of CC events which populate the lowest-lying $1^+$ level will relax through at least one isomeric transition. The ratio of observed intensities for the 448.6~keV and 166.7~keV transitions is 18.2:7, indicating that the first $1^+$ level decays via the transitions $588.8\rightarrow140.3$ and $588.8\rightarrow422.1$ approximately 70\% and 30\% of the time, respectively, subject to a $\sim$5\% uncertainty due to the unknown internal conversion (IC) coefficient of each transition. The former directly feeds the isomeric state at 140.3~keV, guaranteeing that it reaches the ground state through at least one long-lived state. Of the latter, the (208.6~keV):(239.9~keV) intensity ratio indicates that approximately 2\% of decays proceed via the $422.1\rightarrow313.6\rightarrow105.0$ sequence to the ground state composed of purely prompt transitions; the remainder proceed through the isomeric state at 73.7~keV. The branching of decays from the 140.3~keV state remains uncertain. The multipolarity and/or IC coefficients of these low-energy transitions will need to be measured in future work.

The proposed level scheme can be compared to predictions from the SM. 
The model shown in Fig.~\ref{fig:scheme} uses the NuShellX@MSU~\cite{Brown:2014} code with the SN100PN effective interaction~\cite{Brown:2005}, which demonstrates good agreement with prior data~\cite{Astier2013, capetown_placeholder, Wimmer:2011}. 
A key feature of the predicted level structure is the presence of $2^+$ states at 83 and 224~keV, to which the $1^+$ state decays~\cite{Haselschwardt:2020ffr}. Neither has been conclusively identified to date. One candidate is our level at 422.1~keV, which is placed through two-fold coincidences of $108.5$, $239.8$, and $73.7$~keV $\gamma$ rays and which is fed directly by the first $1^+$ state via the $166.7$~keV transition. 
A $2^+$ level at this energy would be unresolved in $^{138}$Ba(d,$\alpha$) reactions from the $(4^+)$ state observed at $423\pm3$~keV~\cite{capetown_placeholder}, which we in turn associate with our reconstructed level at 424.1~keV level due to its strong connection with the $4^+$ state at 105.0~keV via 105.0/319.2~keV coincidences.
Another candidate is the level at 140.2~keV, which is fed directly by the first $1^+$ state via the $588.8\rightarrow140.2\rightarrow73.7\rightarrow{\rm G.S.}$ sequence that is reconstructed by our observed coincidences of the $448.6$, $66.6$, and $73.7$ keV transitions. However, the angular distributions in $^{138}$Ba(d,$\alpha$) reactions favor $J^{\pi} = 3^+$, leaving this interpretation uncertain. Finally, a state at 432(2)~keV, first observed in Ref.~\cite{Frekers:2013} and given a tentative spin-parity assignment of $(3^+)$ therein, was recently identified as a possible $(2^+)$ state in Ref.~\cite{capetown_placeholder}. This state is not reconstructed by any coincidences in the present work, and we do not observe any transitions near 157~keV that would connect this state with the $1^+$ state at 588.8~keV. Thus our data are in tension with the more recent spin-party assignment. Further measurements with both charged-particle and $\gamma$-ray detection will be ideal for fully characterizing this structure.

\emph{Discussion --} 
The isomeric states measured here will produce a unique signature for identification of CC interactions in xenon-based detectors. 
The prompt interaction, composed of the emitted electron and the initial relaxation of the Cs nucleus, will almost always be followed by the delayed emission of one or more $\sim$$100$~keV $\gamma$'s/IC electrons. 
In detectors which measure scintillation this will produce a characteristic two- or three-pulse signal. Based on the lifetimes measured here, an experiment which can resolve these secondary pulses in a 25--1000~ns window after the primary will have $>80\%$ efficiency for tagging CC interactions~\cite{Haselschwardt:2020ffr}. The resolving power will be determined by the time structure of each pulse, which is the convolution of (1) the photoemission of the scintillating medium, (2) photon transport in the detector, and (3) the time resolution of the photon detection system.
In the case of (1), there are three technologies currently being pursued for experiments at the tonne-scale and beyond: LXe TPCs~\cite{NEXO_pCDR_2018,LZ_0nuBB_sensitivity2020,g3_whitepaper}, high-pressure gaseous Xe (GXe) TPCs~\cite{NEXT:2020amj}, and loaded liquid scintillator (LS) detectors~\cite{KamLAND-Zen:2022tow}, which have emission time constants of $\tau_{\rm LXe} = 27$~ns, $\tau_{\rm GXe} = 4$~ns, and $\tau_{\rm LS} = 6$~ns~\cite{Kubota_1978, Akerib:2018kjf, Azevedo:2017egv, Suekane:2004ny}. In modern experiments (2) is $\mathcal{O}(10)$~ns and (3) can be as low as $\mathcal{O}(1)$~ns~\cite{Akerib:2018kjf,AoboThesis}. High efficiencies are therefore possible with existing techniques.

The background from accidental coincidences of scintillation pulses which mimic the energy/time structure of the CC signal are likely to have negligible rate in modern, low-background Xe experiments. For example, an 80~t $^{\rm nat}$Xe detector will have probability $\sim$$10^{-9}$ to accidentally accept coincident 50--100~keV pulses in a 1~$\mu$s window provided its background rate is $\sim$30~evts/(t$\cdot$yr$\cdot$keV) as achieved in XENONnT~\cite{XENONnT_ER}. Further suppression is possible if the spatial information of the accompanying charge signal is used~\cite{Haselschwardt:2020ffr}.

\begin{table}[t]
\centering
\smaller
\caption{Event rates for $^7$Be and CNO solar neutrino capture in current ($^{*}$) and planned experiments which deploy $^{136}$Xe and the previously proposed target isotopes $^{130}$Te and $^{100}$Mo. Rates are given in solar neutrino units (SNU) for each isotope and events per year for each experiment. The mass column indicates the mass of isotope deployed, as opposed to the total payload of the experiment. Scattering rates on $^{130}$Te and $^{100}$Mo are taken from Ref.~\cite{Ejiri:2013jda}, while those on $^{136}$Xe are from Ref.~\cite{Haselschwardt:2020ffr}. } \label{tab:expts}
\begin{ruledtabular}
\begin{tabular}{lcclccc}

 \multirow{2}{*}{Isotope} & \multicolumn{2}{c}{\centering Rate (SNU)}  & \multirow{2}{*}{Experiment} &  Mass & \multicolumn{2}{c}{Rate (evt/yr)} \\
 & $^7$Be & CNO & & (t) & $^7$Be & CNO \\
\hline
\Tstrut \multirow{4}{*}{$^{136}$Xe}  & \multirow{4}{*}{42.5} & \multirow{4}{*}{6.3} & LZ$^{*}$ & 0.62 & 3.7 & 0.54 \\
 & & & KamLAND-Zen$^{*}$ & 0.68 & 4.0 & 0.59 \\
 & & & nEXO &3.2 & 19.0 & 2.8 \\
 & & & 80t $^{\rm nat}$Xe TPC & 7.1 & 42.1 & 6.2 \\[1ex]
$^{130}$Te & 23.2 & 3.7 & SNO+ (0.5\% Te) & 1.3 & 4.4 & 0.70 \\[1ex]
$^{100}$Mo & 126 & 15 & CUPID & 0.25 & 6.1 & 0.72 \\

\end{tabular}
\end{ruledtabular}
\end{table}

Other $\beta\beta$-decay isotopes have been proposed as favorable targets for solar neutrino capture. Suggested schemes for their deployment include metal-loaded LS~\cite{Raghavan:1976yc,Raghavan:1997ad,Wang:2020tdm} or solid-state sensors~\cite{Ejiri:1999rk,Zuber:2002wi}. In Table~\ref{tab:expts}, we compare current and proposed xenon-based experiments with the two most massive experiments of each type planned for the near future\footnote{See e.g. Ref.~\cite{dbd_review}.}: the SNO+~\cite{SNO:2021xpa} and CUPID~\cite{CUPID:2019imh} experiments, which will search for $0\nu\beta\beta$ in $^{130}$Te and $^{100}$Mo, respectively. The solar neutrino CC interaction rates in these next-generation experiments are comparable to those in currently-operating experiments which deploy $^{136}$Xe.

Tagged CC events offer a promising avenue toward improved and complementary measurements of the CNO flux. The present uncertainties reported by Borexino are $^{+30\%}_{-12\%}$~\cite{BOREXINO:2022abl}. 
By comparison, an 80~t $^{\rm nat}$Xe detector may achieve 30\% (12\%) statistical uncertainty in 2 (11) years based on the rates in Table~\ref{tab:expts}.
Larger Xe experiments, proposed for far-future $0\nu\beta\beta$ searches, could reduce the uncertainty by a factor of $\sim$2--7~\cite{Theia:2019non, Avasthi:2021lgy}. 
Uncertainties in the CC interaction cross section (e.g. those from Gamow-Teller strengths~\cite{Frekers:2013}) can be calibrated \emph{in situ} through simultaneous measurement of the $7\times$ stronger $^{7}$Be flux,  which is known to within 3\%~\cite{Borexino:2017rsf}.
Xenon experiments may also combine the tagged CC event sample with an analysis of neutrino-electron elastic scatters for increased sensitivity.

Within the scope of existing experiments, KamLAND-Zen recently reported the largest exposure of $^{136}$Xe to date, totaling 1~t-yr~\cite{KamLAND-Zen:2022tow}. The fast intrinsic timing of the experiment's organic LS is advantageous, though identification of delayed pulses from the 105, 74, or 67~keV transitions will depend on the experiment's energy threshold and the rate of $^{14}$C background in this region. 
The currently operating LZ~\cite{LZ:2019sgr} and XENONnT~\cite{XENON:2020kmp} experiments feature $\sim$1~keV energy thresholds, unprecedentedly low backgrounds, and will amass $^{136}$Xe exposures of $\sim$2~t-yr. Delayed-coincidence searches in any of these experiments could detect roughly 10 solar neutrino events and would provide world-leading sensitivity to MeV-scale fermionic dark matter.

In summary, we have measured the energies and lifetimes of low-lying states in $^{136}$Cs. The isomeric states measured here enable the use of a delayed-coincidence tag to uniquely identify 
CC interactions in experiments which deploy $^{136}$Xe. 
Thus a new channel exists in which to perform novel spectroscopic studies of solar neutrinos and dark matter in current and future xenon-based experiments.

\emph{Acknowledgments --} The authors would like to thank David Radford for helpful discussions and assistance with the RadWare analysis software. We thank Smarajit Triambak for helpful discussions. 
This work was supported by the U.S. Department of Energy Office of Science under contract number DE-AC02-05CH11231, grant no. DE-FG02-97ER41033, and by DOE-NP grant DE-SC0023053. B.~L. was supported by the Department of Energy, Laboratory Directed Research and Development program at SLAC National Accelerator Laboratory, under contract DE-AC02-76F00515 as part of the Panofsky Fellowship.

\FloatBarrier
\bibliographystyle{apsrev4-2}
\bibliography{main}

\end{document}